\documentclass[astrosymb,twocolumn,floatfix]{aastex701}
\usepackage{physics}
\usepackage{xspace}
\usepackage{acro}[=v2]
\acsetup{display-foreign = false}
\acsetup{foreign-format = {\scshape} }
\ExplSyntaxOn
\prop_new:N \l__acro_foreign_format_prop
\ExplSyntaxOff
\usepackage{new_commands}
\date{\today}

\begin{document}
\title{The maximum mass ratio of hierarchical binary black hole mergers may cause the $q$-$\chi_{\rm eff}$ correlation}
\author[0000-0002-4103-0666]{Aditya Vijaykumar}
\email{aditya@utoronto.ca}
\CITA
\author[0000-0002-6121-0285]{Amanda M. Farah}
\email{afarah@cita.utoronto.ca}
\CITA
\author[0000-0002-1980-5293]{Maya Fishbach}
\email{fishbach@cita.utoronto.ca}
\CITA

\begin{abstract}
    Regardless of their initial spins, the merger of two roughly equal mass black holes (BHs) produces a remnant BH of dimensionless spin $0.69$. Such remnants can merge with other BHs in dense stellar environments and produce hierarchical mergers. Analyzing the latest catalog binary black hole (BBH) mergers from the LIGO-Virgo-KAGRA detectors, we identify a subpopulation with primary spins consistent with such hierarchical mergers. Consistent with astrophysical expectations for mergers of second-generation BHs with first-generation BHs, we find that this subpopulation has mass ratios below $0.59^{+0.18}_{-0.23}$. We also infer that $19$-$88\%$ of the BBH population below this mass ratio is consistent with belonging to the hierarchically merged population. Our results offer a natural explanation for the narrowing of the effective inspiral spin distribution with mass ratio observed in other studies.
\end{abstract}


\section{Introduction}
\label{sec:intro}

Under the right conditions, the remnant \acp{BH} of \ac{BBH} mergers can go on to merge with other \acp{BH}. Such mergers are referred to as \textit{hierarchical mergers}, and are a natural outcome of dynamical assembly of \acp{BBH} in dense stellar systems~\citep{2019PhRvD.100d3027R, 2019MNRAS.486.5008A}. Models of dense stellar clusters predict that the most common type of hierarchical mergers involve the merger of a \textit{second-generation} (2G) \ac{BH} with a \textit{first-generation} (1G) \ac{BH}, making up $\approx 10$-$20\%$ of the total mergers from these environments~\citep[e.g.,][]{2020ApJS..247...48K, 2020MNRAS.492.2936A}. \acp{BBH} involving higher-generation \acp{BH} are also predicted, but their rate is suppressed by a factor of at least  $\approx 10$ compared to the 2G+1G rate. Thus, if the total merger rate of \acp{BBH} in dense stellar systems is high, 2G+1G mergers could make up a significant population of systems detected by the  LIGO~\citep{2015CQGra..32g4001L}, Virgo~\citep{2015CQGra..32b4001A} and KAGRA~\citep{2021PTEP.2021eA101A} (LVK) detectors.

Hierarchical mergers have clear population predictions that allow them to be straightforwardly distinguished from the larger \ac{BBH} population \citep{2017ApJ...840L..24F, 2017PhRvD..95l4046G}.
This is in contrast to most other formation pathways whose population predictions suffer from large theoretical uncertainties~\citep{ 2020FrASS...7...38M,2022PhR...955....1M, 2022LRR....25....1M}. 
Hierarchical mergers may therefore be the first subpopulation to be confidently identified, even if they should make up a small fraction of the total population.
Robust predictions of a hierarchically merged subpopulation include the following:
    \begin{enumerate}
        \item Component spin magnitudes peaked near 0.69~\citep{2017ApJ...840L..24F, 2017PhRvD..95l4046G} 
        \item Isotropically distributed spin tilts, on account of merging in dense stellar environments~\citep{2016ApJ...832L...2R}, and 
        \item 2G+1G mergers with mass ratios $q$ below $\approx0.6$ \citep{2017PhRvD..95l4046G,2019PhRvD.100d3027R,2020ApJ...900..177K}.
    \end{enumerate}

The LVK collaboration recently published two \ac{BBH} events with parameters that are compatible with these characteristics~\citep{2025ApJ...993L..21A}.
This motivates the search for a subpopulation of hierarchical mergers on the population level.
Assuming that hierarchical (2G+1G) mergers follow a primary spin-magnitude distribution peaked at $\chi_1 \approx 0.69$ and isotropic spin tilts, we fit for the maximum mass ratio of such a subpopulation. 
We find evidence for the existence of such a subpopulation below mass ratio $q$ of $0.59^{+0.18}_{-0.23}$.
Assuming that this subpopulation does indeed correspond to a population of 2G+1G \acp{BBH}, we measure the fraction and merger rate of hierarchical mergers in the \ac{BBH} population. We also characterize the resultant $q$--$\chi_1$ correlation in the overall \ac{BBH} population and describe how it naturally leads to the observed correlation between effective spin and mass ratio~\citep{2021ApJ...922L...5C, 2025arXiv250818083T}. 

\begin{figure*}
    \centering
    \includegraphics{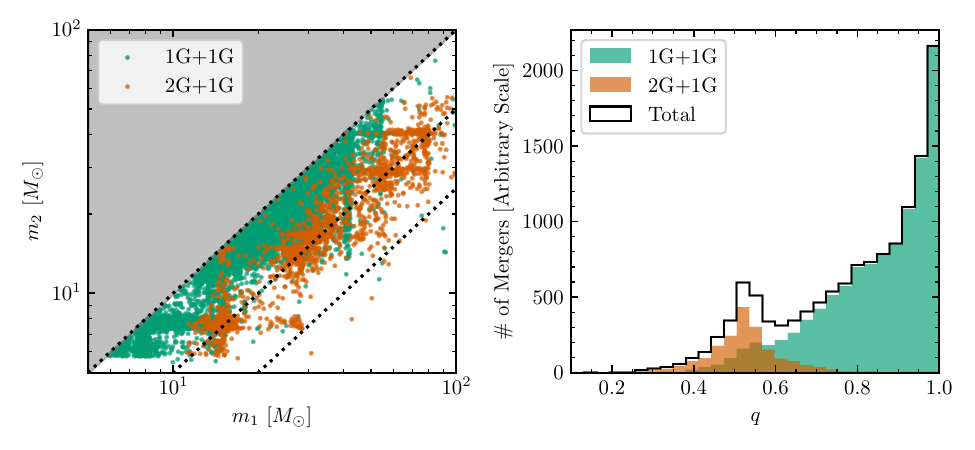}
    \caption{Predictions of 1G+1G and 2G+1G \ac{BBH} mergers from the CMC simulations of dense star clusters~\citep{2020ApJS..247...48K}. The plot on the left shows \ac{BBH} mergers in the $m_1$-$m_2$ plane, where the 2G+1G mergers occupy a region where masses are unequal while 1G+1G mergers occupy the region near equal masses. This is also borne out in the mass ratio distribution of mergers plotted in the right panel, where the mass ratio distribution of 2G+1G mergers peaks at $q\approx 0.5$ whereas that of 1G+1G mergers peaks at $q\approx 1$.}
    \label{fig:CMC-2G}
\end{figure*}

This Letter is structured as follows. In Section~\ref{sec:model-summary}, we provide a description of the phenomenological model used in the work, along with the astrophysical reasoning behind our parameterization. Section~\ref{sec:results} describes results from our population fits, and Section~\ref{sec:q-chieff} describes the implications of our results for the correlation between mass ratio and effective spin. We end with a summary and discussion in Section~\ref{sec:conclusion}. The appendices contain a detailed description of our model along with result validation studies and additional figures.  Throughout the Letter, we use the Planck 2015 cosmology~\citep{2016A&A...594A..13P} as defined in \textsc{LAL}~\citep{lalsuite}. All results are quoted as $90\%$ credible intervals.

\section{Model Summary}
\label{sec:model-summary}
We wish to write down a model that exploits the robustness of predictions for hierarchical mergers as an anchor to search for such a subpopulation in the data. In what follows, we will describe the spin magnitude, spin tilt, and mass ratio expectations for hierarchical mergers and use these to motivate a model to search for them. Although our model is motivated by theoretical arguments, it does allow for the \textit{absence} of a hierarchical subpopulation. As we will show in Section~\ref{sec:results}, we find consistent evidence for such a subpopulation in the data. 

\subsection{Astrophysical motivation}
\label{subsec:background}

When two \acp{BH} merge, angular momentum conservation dictates that the final remnant \ac{BH} acquires the combined spin angular momenta of the progenitor \acp{BH} and their orbital angular momentum, after accounting for the angular momentum carried away by \ac{GW} emission during the merger process. While a rough estimate of the final spin can be obtained using a Newtonian treatment~\citep[e.g.][]{2008PhRvD..77b6004B}, precise estimates of the remnant spin are only possible through numerical simulations~\citep[e.g.,][]{2009PhRvD..79b4003S} and fitting formulae built from them~\citep[e.g.,][]{2019PhRvL.122a1101V}. One robust prediction is that the remnant spins of \acp{BBH} having $q\gtrsim0.5$ peak near 
of ${\approx}0.69$~\citep{2008ApJ...684..822B, 2016ApJ...825L..19H, 2017ApJ...840L..24F, 2017PhRvD..95l4046G, 2019PhRvD.100d3027R, 2021NatAs...5..749G} regardless of the initial spin distribution.

In dense stellar clusters, \acp{BBH} are assembled through dynamical interactions rather than processes involving isolated binary stars. Due to the random nature of these interactions, they break any initially existing symmetry or preferential alignment between the spin angular momenta of the black holes and the orbital angular momentum. Thus, it is expected that the spin directions of \textit{all} \acp{BBH} in a cluster, including those that are hierarchically merged,  would be oriented isotropically, i.e., would have no preferred direction~\citep{2016ApJ...832L...2R}\footnote{See \citet{2025ApJ...983L...9K} for one example of when \acp{BBH} in star clusters can retain preferential alignment.}. 

Another process that is dynamically important in dense star clusters is mass segregation. Mass segregation drives the more massive \acp{BH} toward the cluster core. Owing to the high central density, these \acp{BH} experience frequent dynamical encounters, typically forming binaries that merge with mass ratios close to unity ($q \approx 1$) for 1G+1G mergers~\citep[e.g.,][]{1987degc.book.....S, 2003gmbp.book.....H, 2018PhRvL.120o1101R}. If the recoil kick imparted to the remnant 2G \ac{BH} is smaller than the cluster’s escape velocity, the remnant remains bound near the center. Having formed from two of the most massive \acp{BH} in the cluster, the 2G \ac{BH} becomes the most massive object in its vicinity and is therefore the most dynamically active and is likely to merge with a 1G \ac{BH} before another 2G \ac{BH} is formed. Thus, the resultant mass ratio of the mergers will lie around $q\approx0.5$~\citep{2019PhRvD.100d3027R, 2019MNRAS.486.5008A, 2026ApJ...997..267Y}. The mass ratio distribution could be different in more massive clusters where 1G+1G merger timescales are shorter~\citep{2019PhRvD.100d3027R}.

\begin{figure*}[!htbp]
    \centering
    \includegraphics{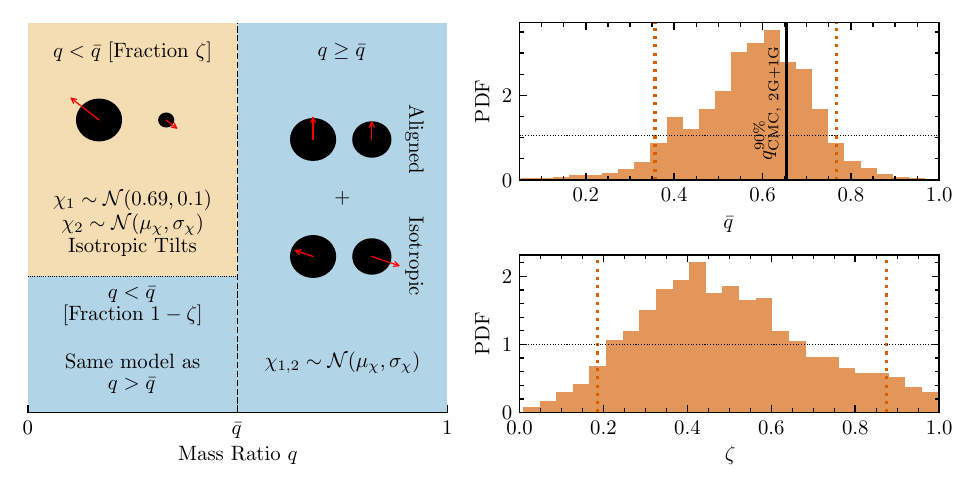}
    \caption{The left panel shows a visual summary of how our phenomenological model is constructed and parameterized. The mass ratio scale $\bar{q}$ forms a threshold below which a fraction $\zeta$ of the population is composed of hierarchical (2G+1G) mergers. The two panels on the right show the measurements of $\bar{q}$ and $\zeta$ using the latest catalog of \ac{GW} signals. In both these panels, the horizontal gray dotted lines denote the prior on each parameter, whereas the vertical red dotted lines denote the 90\% credible region of the posterior. We obtain $\bar{q} = 0.59^{+0.18}_{-0.23}$, with $\zeta= 19$-$88$\% of the population below $\bar{q}$ belonging to the hierarchically merged population. For reference, we also plot the 90\% upper limit of the 2G+1G mass ratio distribution $q^{90\%}_\mathrm{CMC, 2G+1G} = 0.65$ from the CMC simulations as a vertical black solid line. }
    \label{fig:infographic_and_constraints}
\end{figure*}

To illustrate this, we plot the dataset of \ac{BBH} mergers from the Cluster Monte Carlo (CMC) catalog~\citep{2020ApJS..247...48K} split by their generation in Figure~\ref{fig:CMC-2G}. We also plot the histogram of mass ratios of these mergers. As is evident, most 2G+1G \acp{BH} prefer having unequal masses, with their mass ratio distribution peaking around $q\approx 0.5$. Another trend worth noting is that 1G+1G \acp{BH} dominate the total observed population above $q\gtrsim0.6$. Thus, if we want to search for a 2G+1G population in the data, one way would be to look for a transition mass ratio below which the spin distribution has at least some support for primary spins around 0.69 and isotropic spin tilts.

\subsection{Model}
\label{subsec:model}
To investigate the features described in the preceding section, we employ a phenomenological population model. For the primary mass distribution, we adopt the \textsc{Broken Power Law + Two Peaks} mass model used in the \citet{2025arXiv250818083T}, but also allowing for a maximum secondary mass as done in \citet{2025arXiv250904151T}. We introduce a mass ratio threshold $\bar{q}$ that serves as a sharp boundary separating two distinct subpopulations with independent spin distributions. For binaries with $q \ge \bar{q}$, we utilize the spin distribution models from \citet{2025arXiv250818083T}, parameterizing spin magnitudes as a normal distribution with mean $\mu_\chi$ and width $\sigma_\chi$ truncated to the physical range $[0, 1]$. Spin tilts in this regime are allowed to have an isotropic component as well as a component parameterizing preferential alignment. For $q < \bar{q}$, we posit that a fraction $\zeta$ of the population follows an isotropic spin distribution with primary spin magnitudes consistent with hierarchical mergers, i.e., a truncated normal distribution with mean 0.69 and width 0.1. While this is a strong prior choice, we show in Appendix~\ref{appendix:comparison-GW241011} that varying the width does not affect the inference from the model significantly. The secondary spin distribution follows the spin magnitudes in the main population. The rest of the mergers (fraction $1-\zeta$) in the $q <\bar{q}$ population follow the same spin distribution as the $q \ge \bar{q}$ population.  

A schematic visualization of the model is provided in the left panel of Figure~\ref{fig:infographic_and_constraints}, and a more detailed description of the model can be found in Appendix~\ref{appendix:model_details}.



\begin{figure}
    \centering
\includegraphics{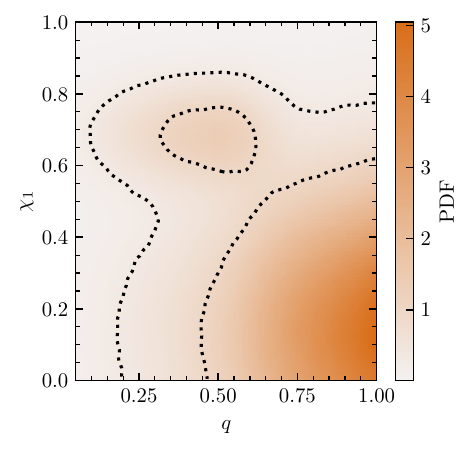}
    \caption{Smoothed histogram of fair draws from the inferred population in the spin-magnitude--mass ratio plane. Two clear regions emerge in this plot: a lower-spin population concentrated toward $q\approx 1$, as well as a $\chi_1 \approx 0.7$ population concentrated around $q \approx 0.5$. Thus, there is an apparent anticorrelation between the primary spin magnitude and mass ratio.}
    \label{fig:q-chi1}
\end{figure}

\section{Results}
\label{sec:results}
Using a hierarchical Bayesian inference framework~\citep{2004AIPC..735..195L,2019PASA...36...10T,2022hgwa.bookE..45V}, we fit the model described in Section~\ref{subsec:model} to data from \ac{GWTC-4.0}. We use the same dataset as used in the LVK Collaboration's analysis of \ac{GWTC-4.0}~\citep{2025arXiv250818083T}, but with two changes:
\begin{enumerate}
    \item We exclude the massive event GW231123~\citep{2025ApJ...993L..25A} from the dataset.
    \item We include the events GW241011 and GW241110~\citep{2025ApJ...993L..21A} released from the second half of observing run 4 (O4). These events have been proposed to be consistent with hierarchical origin~\citep{2025ApJ...993L..21A}.
\end{enumerate}
While this forms our default dataset, our results do not change significantly either by adding GW231123 or leaving out GW241011 and GW241110. Consistency checks leaving out the latter can be found in Appendix~\ref{appendix:comparison-GW241011}. To model the sensitivity of the \ac{GW} detectors and correct for selection effects, we use the injection campaign from \citet{2025PhRvD.112j2001E}. We use the \textsc{GWPopulation}~\citep{2025JOSS...10.7753T}, \textsc{GWPopulation\_pipe}~\citep{2021zndo...5654673T}, and \textsc{Bilby}~ software packages for our inference, sampling the posterior with the nested sampler implemented in the \textsc{Dynesty} software package~\citep{2020MNRAS.493.3132S}.

The main population parameters of interest are $\bar{q}$ and $\zeta$. We find that a model with a transition in spins at $\bar{q}$ is preferred over a model where there is no such transition by a Bayes factor of $\approx 15$. The right panel of Figure~\ref{fig:infographic_and_constraints} shows the results from this model. We measure $\bar{q} = 0.59^{+0.18}_{-0.23}$, implying that the putative hierarchically merged population prefers asymmetric masses below this mass ratio scale. This result is consistent with expectations from the CMC simulation suite, which predicts a 90\% upper limit of $q=0.65$ for the 2G+1G population averaged across cluster masses, metallicities, and virial radii. Furthermore, we infer that $\zeta = 19$-$88\%$ (90\% credible interval) of the population below $\bar{q}$ is composed of the putative hierarchically merged subpopulation, corresponding to a merger rate at redshift $z=0.2$ of $1.6^{+3.3}_{-1.4} \, \mathrm{Gpc}^{-3} \, \mathrm{yr}^{-1}$. This inferred merger rate is consistent with the lower limit of the 2G+1G merger rate inferred in \citet{2025arXiv250904151T}, as well as the 2G+1G merger rates inferred \citet{2025arXiv251105316T} and \citet{2026arXiv260103456F}. In Appendix~\ref{appendix:comparison-GW241011}, we examine the dependence of these results on the \added{Monte Carlo uncertainty thresholds~\citep{2022arXiv220400461E, 2023MNRAS.526.3495T,2025arXiv250907221H} applied to our likelihood calculations.}

 Assuming a scaling of $\approx 7.5$ between the 2G+1G rate and the total merger rate in dense clusters (i.e. 1G+1G and all hierarchical mergers), consistent with predictions from CMC simulations~\citep[e.g.][]{2026ApJ...998..138M}, we find that a median of 40\% of all binaries merge in dense star clusters. However, this quantity is not constrained precisely, with the 90\% credible  region being $5$-$100\%$.

 \begin{figure*}
    \centering
    \includegraphics{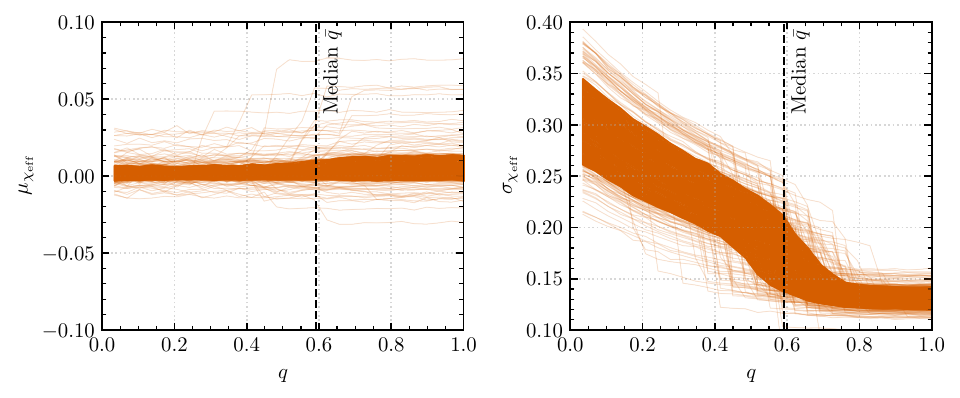}
    \caption{Mean $\mu_{\chi_\mathrm{eff}}$ and width $\sigma_{\chi_\mathrm{eff}}$ of the $\chi_\mathrm{eff}$ distribution inferred from our model as a function of mass ratio $q$. The light red lines denote traces from different posterior samples, whereas the shaded region is the inferred 90\% credible region. While $\mu_{\chi_\mathrm{eff}}$ does not vary significantly with $q$, $\sigma_{\chi_\mathrm{eff}}$ transitions from a plateau at $q\gtrsim0.7$ to a linear curve with negative slope at $q\lesssim 0.6$. Thus, the width of the $\chi_\mathrm{eff}$ distribution decreases with increasing mass ratio.}
    \label{fig:q-chieff}
\end{figure*}

To further visualize our results, we plot the density of $10^7$ fair draws from the population in the $q$--$\chi_1$ plane in Figure~\ref{fig:q-chi1}. In this plane, the separation between the two populations is clearly visible. Near $q\approx 1$, most mergers have low spins, with many having spins consistent with zero. On the other hand, the region with $q \lesssim 0.6$ is dominated by primary spins centered around 0.69. This effectively showcases the correlation in the $q$--$\chi_1$ plane---mergers with highly spinning primaries consistent with hierarchical mergers prefer to lie at unequal masses when compared to those inconsistent with hierarchical mergers. 

\section{Consequences for the correlation between effective inspiral spin and mass ratio}
\label{sec:q-chieff}

The effective inspiral spin $\chi_\mathrm{eff}$ is defined as the mass-weighted sum of component spin vectors $\vec{\chi}_1$ and $\vec{\chi}_2$ aligned along the direction of the orbital angular momentum $\vec{L}_\mathrm{orb}$. That is,
\begin{equation}
    \chi_\mathrm{eff} = \dfrac{m_1 \vec{\chi}_1 + m_2 \vec{\chi}_2} {m_1 + m_2} \vdot \vec{L}_\mathrm{orb} \ .
\end{equation}
The effective inspiral spin is approximately conserved during the inspiral of a \ac{BBH}, and it encodes the leading-order dependence of the component spins on the orbit  and consequently on the \ac{GW} signal emanating from the system~\citep{2001PhRvD..64l4013D, 2008PhRvD..78d4021R, 2011PhRvL.106x1101A}. At the population level, a correlation between the $q$ and $\chi_\mathrm{eff}$ has been known since the work of \citet{2021ApJ...922L...5C} who found that, when the $\chi_\mathrm{eff}$ distribution is modeled as a Gaussian at each $q$ value, the \textit{mean} of the $\chi_\mathrm{eff}$ distribution decreases with increasing $q$. More recently, it was found that either the \textit{mean} or the \textit{width} of the $\chi_\mathrm{eff}$ distribution can vary as a function of $q$~\citep{2025arXiv250818083T}, with there being preference for the distribution narrowing as a function of $q$. We find that this latter result follows naturally from our inference of the $q$--$\chi_1$ correlation.

For each posterior sample of our population parameters, we generate $10^4$ fair draws from the population of masses and spins. \added{We convert the spins to $\chi_\mathrm{eff}$ and plot the mean $\mu_{\chi_\mathrm{eff}}$ and width (standard deviation) $\sigma_{\chi_\mathrm{eff}}$ of the $\chi_\mathrm{eff}$ samples as a function of mass ratio.} We do not make the assumption that the $\chi_\mathrm{eff}$ distribution is Gaussian distributed in each $ q$ bin---in fact, this is not generally true in our model at all values of $q$. The results are plotted in Figure~\ref{fig:q-chieff}. The left panel shows $\mu_{\chi_\mathrm{eff}}$ as a function of mass ratio. This quantity is not visibly correlated with the mass ratio, and hovers around zero across the entire range of $q$.  On the other hand, $\sigma_{\chi_\mathrm{eff}}$ is correlated with $q$---the width of the $\chi_{\mathrm{eff}}$ distribution narrows as a function of $q$. \added{The median $\sigma_{\chi_\mathrm{eff}}$--$q$ correlation is not linear throughout the entire mass ratio space, but rather is composed of a plateau above $q\approx 0.7$, a linear regime with slope of ${\approx}{-0.2}$ below $q\approx0.6$, and a nonlinear transition in the middle.} This correlation is a direct consequence of the $q$-$\chi_1$ anticorrelation in Figure~\ref{fig:q-chi1}, coupled with the isotropic spin distribution assumed for the low-$q$ population. Taken at face value, these results suggest that using a linear model for correlations between the mean and width of the $\chi_\mathrm{eff}$ distribution and $q$ perhaps amounts to model misspecification, and future analyses should concentrate on more flexible parametric models or nonparametric models~\citep{2022MNRAS.517.3928A, 2023ApJ...958...13A, 2025PhRvD.111f3043H, 2025ApJ...991...17R} to capture the correlation. Due to the differences in assumptions made in the model here and the ones used in \citet{2025arXiv250818083T}, we do not expect the inference to match perfectly between the two. Nevertheless, as we show in Appendix~\ref{appendix:comparison-GW241011}, the results are qualitatively consistent.


\section{Discussion}
\label{sec:conclusion}
In this Letter, we used robust predictions for the spins of hierarchically merged 2G \acp{BH} to characterize their subpopulation in the distribution of \ac{BBH} masses and spins. We found that such a subpopulation lies below a mass ratio of $\bar{q} = 0.59^{+0.18}_{-0.23}$, and could compose $19$-$88\%$ of the population in this region. Our work adds to the growing evidence of hierarchical mergers in \ac{GW} data~\citep{2020ApJ...900..177K, 2021ApJ...915L..35K, 
2024arXiv240601679P,
2024PhRvL.133e1401L,
2025ApJ...987...65L,
2025arXiv250904637A,
2025PhRvL.134a1401A, 2025arXiv251105316T,
2025arXiv250915646B, 2025arXiv251203152B, 2025arXiv250909123A,2026arXiv260103456F} and provides estimates of the 2G+1G merger rates consistent with these works. These results imply the existence of an anticorrelation in the $q$--$\chi_1$ plane and naturally predict the narrowing of the $\chi_\mathrm{eff}$ distribution with mass ratio. The approach used here also does not explicitly model the primary mass distribution of hierarchical mergers differently from rest of the population, although e.g., 2G+1G mergers past $m_2\approx 45\,\Msun$ cutoff \citep{2025arXiv250904151T,2025arXiv250904637A} are implicitly accounted for through the $q$-dependence of the $\chi_1$ distribution. In a companion Letter~\citep{2026arXiv260103456F}, we account for different mass distributions for these subpopulations and show that 2G+1G mergers likely have a merger rate that evolves more steeply with redshift as compared to the rest of the population, thereby providing a potential explanation for the observed broadening of the $\chi_\mathrm{eff}$ distribution with redshift~\citep{2022ApJ...932L..19B, 2025arXiv250818083T}. Taken together with results of this work, a promising picture involving hierarchical mergers to explain various observed correlations in the \ac{BBH} population is emerging. While we use a strongly modeled approach for the BBH population in this work, it should be straightforward to verify our results using the many weakly modeled approaches suggested in the literature~\citep{2023arXiv230401288G, 2025PhRvD.111f3043H, 2025ApJ...991...17R, 2025arXiv251025579T}. 

Several previous studies have also searched for subpopulations of hierarchical mergers using phenomenological models.
Many of these make use of more sophisticated and self-consistent models \citep[e.g.][]{2020ApJ...893...35D,2020ApJ...900..177K,2021ApJ...915L..35K, 2022PhRvD.106j3013M, 2022ApJ...941L..39W}.
Our approach is different as it chooses to anchor the hierarchical subpopulation to a single and highly robust theoretical prediction. In spirit, our approach is most similar to that presented in \cite{2023ApJ...946...50B}, which looks for separate dynamical and isolated channels through spin orientations but does not explicitly apply their methodology to search for hierarchical mergers.
\added{While we assume a fixed, isotropic spin tilt distribution for the hierarchical subpopulation, jointly inferring the spin tilt distribution yields results consistent with isotropy, in agreement with \citet{2026arXiv260103456F}. The resulting impact on the $\bar{q}$ and $\zeta$ posteriors is negligible.}
We also note that some studies~\citep{2025arXiv250923897L, 2026ApJ...996...71H} identify an aligned, highly spinning subpopulation of \acp{BBH}, which has been proposed as an explanation for the observed variation in the mean of the $\chi_\mathrm{eff}$ distribution with $q$~\citep{2025arXiv250923897L}. However, this subpopulation is characterized predominantly by spin magnitudes $\gtrsim 0.8$ and is not explicitly targeted by the population model in this work.

The methods used in this Letter and its companion~\citep{2026arXiv260103456F} essentially isolate the hierarchically merged subpopulation from the rest of the population. Assuming that these hierarchical mergers take place in dense star clusters, one can use these to learn about their properties, especially at high redshift where one cannot access these clusters electromagnetically. For example, the \ac{BBH} mass distribution and evolution of the merger rate with redshift can be used to study the mass, radius, escape velocity, and metallicity distribution of dense clusters in the Universe~\citep{2023MNRAS.522.5546F}. Furthermore, a robust prediction from models of dense star clusters is the existence of eccentric \ac{BBH} mergers~\citep[e.g.][]{2018PhRvL.120o1101R,2021ApJ...921L..43Z, 2024ApJ...969..132V} making up a few per cent of the total population from dense star clusters. This suggests that the hints of eccentric mergers~\citep[e.g.][]{2022ApJ...940..171R, 2025PhRvD.112j4045G} found in \ac{GW} data are not surprising. 

Looking ahead, data from the rest of LVK's fourth observing run is expected to increase the total number of \acp{BBH} by a factor of 2~\citep{2018LRR....21....3A}. This will allow for a deeper characterization of the subpopulation of 2G+1G mergers and more precise constraints on the merger rate of such systems.

\begin{acknowledgements}
We would like to thank Sourav Chatterjee, Thomas Dent, Ish Gupta, Daniel Holz, Kyle Kremer, Utkarsh Mali, Gregoire Pierra, Hui Tong, Claire Ye, and Mike Zevin for many useful conversations. We extend special thanks to Bart Ripperda for granting us access to the \textsc{Bee} workstation at CITA, where the computations for this work were performed.

AV acknowledges support from the Natural Sciences and Engineering Research Council of Canada (NSERC) (funding reference number 568580). MF acknowledges support from the NSERC under grant RGPIN-2023-05511, the Alfred P. Sloan Foundation, and the Ontario Early Researcher Award.
This material is based upon work supported by NSF’s
LIGO Laboratory which is a major facility fully funded
by the National Science Foundation.
This research was supported in part by grant NSF PHY-2309135 to the Kavli Institute for Theoretical Physics (KITP), and has made use of the Astrophysics Data System, funded by NASA under Cooperative Agreement 80NSSC21M00561.

\end{acknowledgements}

\begin{software}
    \ This work made use of the following software packages: \texttt{astropy} \citep{astropy:2013,astropy:2018,astropy:2022}, \texttt{Jupyter} \citep{2007CSE.....9c..21P,kluyver2016jupyter}, \texttt{matplotlib} \citep{Hunter:2007}, \texttt{numpy} \citep{numpy}, \texttt{pandas} \citep{mckinney-proc-scipy-2010,pandas_17992932}, \texttt{python} \citep{python}, \texttt{scipy} \citep{2020SciPy-NMeth,scipy_17873309}, \texttt{Bilby} \citep{bilby_paper,bilby_paper_2,Bilby_17533961}, \texttt{corner.py} \citep{corner-Foreman-Mackey-2016,corner.py_14209694}, \texttt{Cython} \citep{cython:2011}, \texttt{JAX} \citep{jax2018github}, and \texttt{tqdm} \citep{tqdm_14231923},  \texttt{Dynesty} \citep{2020MNRAS.493.3132S}, \texttt{gwpopulation} \citep{2025JOSS...10.7753T}, \texttt{gwpopulation\_pipe} \citep{2021zndo...5654673T}.

    Software citation information aggregated using \texttt{\href{https://www.tomwagg.com/software-citation-station/}{The Software Citation Station}} \citep{software-citation-station-paper,software-citation-station-zenodo}.
\end{software}
\appendix

\section{Model Details}
\label{appendix:model_details}
To model the \ac{BBH} population, we use the \textsc{Broken Power Law + Two Peaks} (BP2P) as our starting point. This model and its associated parameters $\Lambda$ are described in the Appendix of \citet{2025arXiv250818083T}. Following \citet{2025arXiv250904151T}, we modify the mass ratio $q$ distribution to account for a cutoff/gap $m_\mathrm{thresh}$ in secondary masses. The mass model can be written as,
\begin{equation}
    \pi(m_1, q | \Lambda) = \mathrm{BP2P}(m_1 | \Lambda) \ \pi(q | m_1, \Lambda) \qq{;} \pi(q | m_1, \Lambda) = \begin{cases}
        0 \qq{if} m_1 q > m_\mathrm{thresh} \\
        q^{\beta_q} \qq{otherwise} .
    \end{cases}
\end{equation}

To parameterize the distribution of spin magnitudes and cosine of the tilt angles ($\cos \theta_1$ and $\cos \theta_2$ for the primary and secondary respectively), we define a mass ratio scale $\bar{q}$ that separates the hierarchical subpopulation from the rest of the population.

\begin{align}
    \pi(\chi_1, \chi_2, \cos\theta_1,\cos\theta_2 | q, \Lambda) &= \Theta[q < \bar{q}]\qty{\zeta \dfrac{\mathcal{N}^{\chi_1}_{[0,1]}(0.69, 0.1) \mathcal{N}^{\chi_2}_{[0,1]}(\mu_\chi, \sigma_\chi)}{4} + (1 - \zeta) \pi_\mathrm{default}(\chi_1, \chi_2, \cos\theta_1,\cos\theta_2 | \Lambda)} \\
    &+ \Theta[q \ge \bar{q}] \qty{ \pi_\mathrm{default}(\chi_1, \chi_2, \cos\theta_1,\cos\theta_2 | \Lambda)}. 
\end{align}
Here, $\pi_\mathrm{default}$ is given by,
\begin{equation}
    \pi_\mathrm{default}(\chi_1, \chi_2, \cos\theta_1,\cos\theta_2 | \Lambda) = \mathcal{N}^{\chi_1}_{[0,1]}(\mu_\chi, \sigma_\chi) \mathcal{N}^{\chi_2}_{[0,1]}(\mu_\chi, \sigma_\chi) \qty{\dfrac{\xi}{4} + (1 - \xi) \mathcal{N}^{\cos\theta_1}_{[-1, 1]}(\mu_t, \sigma_t) \mathcal{N}^{\cos\theta_2}_{[-1, 1]}(\mu_t, \sigma_t)}
\end{equation}
where $\mathcal{N}^{x}_{[x_1,x_2]}(\mu, \sigma)$ denotes a normal distribution in $x$ with mean $\mu$ and standard deviation $\sigma$ truncated between $x_1$ and $x_2$. The full model can be written as,
\begin{equation}
    \pi(m_1, q, \chi_1, \chi_2, \cos\theta_1,\cos\theta_2 | \Lambda) = \pi(m_1, q | \Lambda) \ \pi(\chi_1, \chi_2, \cos\theta_1,\cos\theta_2 | q, \Lambda) .
\end{equation}

\section{Validation Studies and Additional Figures}
\label{appendix:comparison-GW241011}

In this section, we present a set of validation studies examining the sensitivity of our results to the analysis choices adopted in this work, along with additional supporting figures.

\begin{figure}
    \centering
    \includegraphics{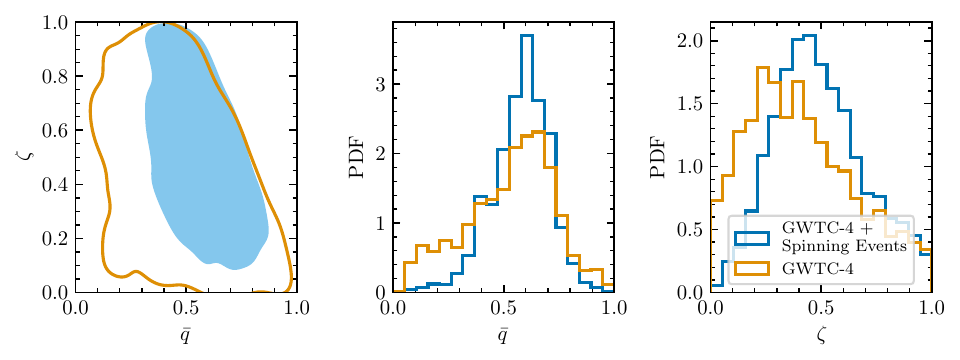}
    \caption{Comparison of $\bar{q}$ and $\zeta$ posteriors obtained by including and excluding GW241011 and GW241110 (referred to as ``Spinning Events'' in the figure legend) in the population. For all results in the main text, we include GW241011 and GW241110 in our dataset. \added{The left panel shows the 90\% credible regions in the $\bar{q}$-$\zeta$ space, while the other two panels show the marginal posteriors}}
    \label{fig:robustness}
\end{figure}

The posterior PDFs of $\bar{q}$ and $\zeta$ obtained with and without the inclusion of the spinning, asymmetric, and misaligned \ac{BBH} mergers GW241011 and GW241110~\citep{2025ApJ...993L..21A} are shown in Figure~\ref{fig:robustness}. As expected, the constraints on $\bar{q}$ and $\zeta$ improve when these events are included. Notably, a non-trivial measurement of $\bar{q}$ persists even when these events are excluded, indicating that the trend reported in this Letter is already present in the \ac{GWTC-4.0} data. The shift of the posterior peak of $\zeta$ toward higher values upon the inclusion of GW241011 and GW241110 is also expected, as we do not self-consistently account for the sensitivity of the second half of O4 in our analysis. Finally, the apparent railing against $\zeta = 0$ in the \ac{GWTC-4.0}–only posterior is driven by its correlation with low values of $\bar{q}$.

\begin{figure}
    \centering
    \includegraphics{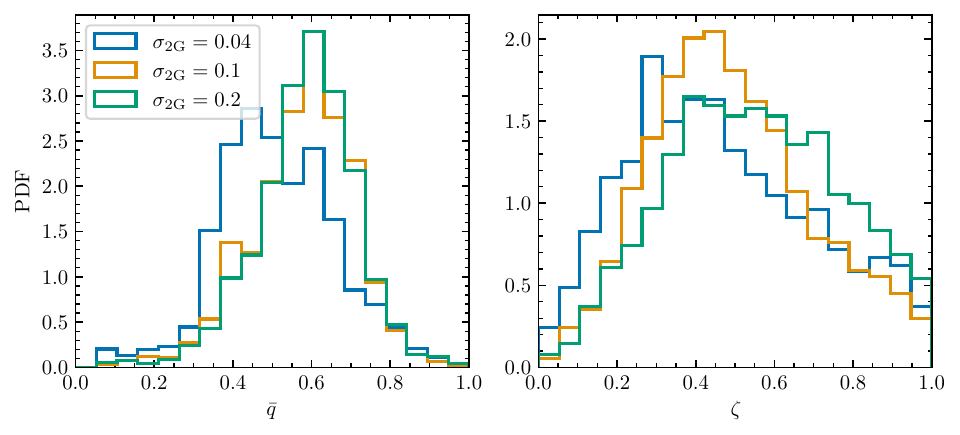}
    \caption{Comparison of $\bar{q}$ and $\zeta$ posteriors obtained by varying widths $\sigma_\mathrm{2G}$ of the 2G spin distribution. The mean of the spin distribution is fixed to 0.69. For results in the main text, we fix $\sigma_\mathrm{2G} = 0.1$.}
    \label{fig:robustness-against-sigma}
\end{figure}

We also present results obtained by varying the width, $\sigma_\mathrm{2G}$, of the 2G spin distribution in Figure~\ref{fig:robustness-against-sigma}. Our fiducial analysis assumes $\sigma_\mathrm{2G} = 0.1$, with the corresponding mean fixed at 0.69. We find that the posterior distributions of $\bar{q}$ and $\zeta$ are largely insensitive to changes in the width of the 2G spin distribution.

\begin{figure}
    \centering
    \includegraphics{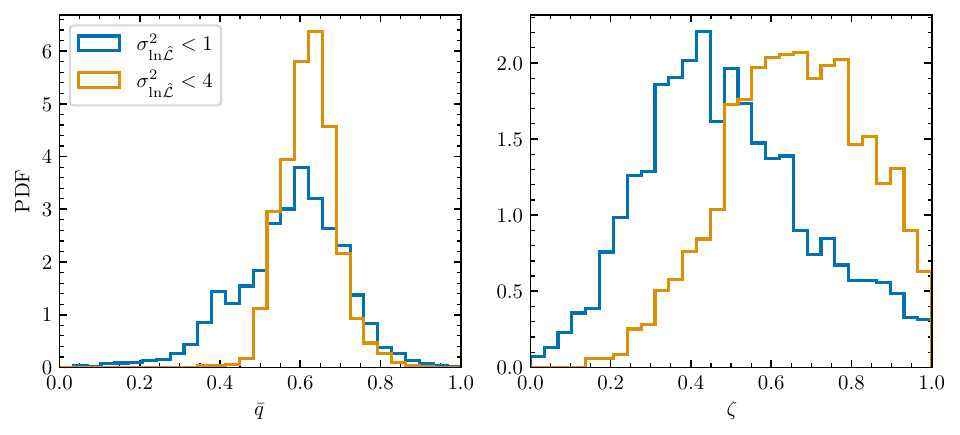}
    \caption{Comparison of $\bar{q}$ and $\zeta$ posteriors obtained by varying the log-likelihood variance threshold. The posteriors become sharper when the default variance threshold $\sigma^2_{\mathrm{ln} \hat{\mathcal{L}}}(\Lambda) < 1$ is relaxed to $\sigma^2_{\mathrm{ln} \hat{\mathcal{L}}}(\Lambda) <4$. We use the conservative choice of $\sigma^2_{\mathrm{ln} \hat{\mathcal{L}}}(\Lambda) < 1$ for all results presented in the main text.}
    \label{fig:robustness-against-maxunc}
\end{figure}

For the default results presented in the main text, we require the log-likelihood variance to satisfy $\sigma^2_{\mathrm{ln}\hat{\mathcal{L}}}(\Lambda) < 1$ \citep[e.g.,][]{2022arXiv220400461E, 2023MNRAS.526.3495T}. This requirement is enforced by setting the log-likelihood to $-\infty$ whenever the condition is not met. When this criterion is relaxed to $\sigma^2_{\mathrm{ln}\hat{\mathcal{L}}}(\Lambda) < 4$, the posterior constraints on $\bar{q}$ and $\zeta$ tighten, as shown in Figure~\ref{fig:robustness-against-maxunc}. Nevertheless, for the results presented in the main text, we adopt the more conservative choice of $\sigma^2_{\mathrm{ln}\hat{\mathcal{L}}}(\Lambda) < 1$.

\begin{figure}
    \centering
    \includegraphics{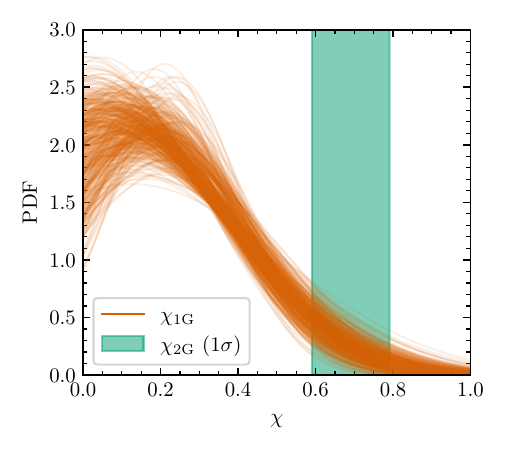}
    \caption{Inferred spin magnitude distribution for the putative 1G \acp{BH} plotted in red traces. The green shaded region encodes the 1$\sigma$ (68\%) region of the 2G spin distribution (normal distribution with mean $0.69$ and width $0.1$ truncated between $0$ and $1$).}
    \label{fig:inferred_chi_distribution}
\end{figure}

Figure~\ref{fig:inferred_chi_distribution} shows the inferred putative 1G spin-magnitude distribution, together with the fixed 2G spin-magnitude distribution adopted in this work. While the 1G distribution peaks at low spin magnitudes, it exhibits a tail extending to spin magnitudes within the range assumed for the 2G population.

\begin{figure}
    \centering
    \includegraphics{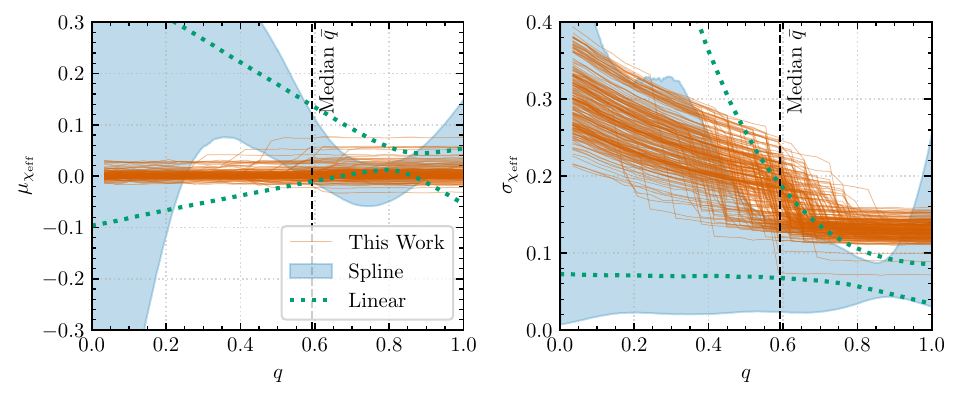}
    \caption{Similar to Figure~\ref{fig:q-chieff} but comparing the mean and width of the $\chi_\mathrm{eff}$ distribution as a function of $q$ inferred with the model in this work to those inferred in \citet{2025arXiv250818083T} using the Linear and Spline models. Although precise agreement is not expected given the differences between these models, the inferred widths are consistent especially at $\mu \lesssim 0.6$.}
    \label{fig:chi_eff_q_lvk}
\end{figure}

Figure~\ref{fig:chi_eff_q_lvk} compares the mean and width of the $\chi_\mathrm{eff}$ distribution inferred as a function of $q$ in this work to those inferred in \citet{2025arXiv250818083T}. The trends are consistent between the model used here and in \citet{2025arXiv250818083T}, although quantitative details differ.

\bibliography{references,NonADSReferences}
\bibliographystyle{aasjournalv7}
\end{document}